\documentclass[aps,prl,showpacs,floatfix,twocolumn,superscriptaddress]{revtex4-1}
\usepackage{graphicx}
\usepackage{bm,amsmath}
\usepackage{dcolumn}
\usepackage{amsfonts,amssymb}
\usepackage{chemfig}
\usepackage{diagbox}

\begin{document}

\title{Monte Carlo Renormalization Flows in the Space of Relevant and Irrelevant Operators:
       Application to Three-Dimensional Clock Models}

\author{Hui Shao}
\email{huishao@bnu.edu.cn}
\affiliation{Center for Advanced Quantum Studies, Department of Physics, Beijing Normal University, Beijing 100875, China}
\affiliation{Beijing Computational Science Research Center, Beijing 100193, China}

\author{Wenan Guo}
\email{waguo@bnu.edu.cn}
\affiliation{Department of Physics, Beijing Normal University, Beijing 100875, China}
\affiliation{Beijing Computational Science Research Center, Beijing 100193, China}

\author{Anders W. Sandvik}
\email{sandvik@bu.edu}
\affiliation{Department of Physics, Boston University, 590 Commonwealth Avenue, Boston, Massachusetts 02215, USA}
\affiliation{Beijing National Laboratory for Condensed Matter Physics and Institute of Physics, Chinese Academy of Sciences, Beijing 100190, China}
\affiliation{Beijing Computational Science Research Center, Beijing 100193, China}

\date{\today}

\begin{abstract}
We study renormalization group flows in a space of observables computed by Monte Carlo simulations. As an example, we consider 
three-dimensional clock models, i.e., the XY spin model perturbed by a $Z_q$ symmetric anisotropy field. For $q=4,5,6$, a scaling 
function with two relevant arguments describes  all stages of the complex renormalization flow at the critical point and in the 
ordered phase, including the cross-over from the U(1) Nambu-Goldstone fixed point to the ultimate $Z_q$ symmetry-breaking fixed point. 
We expect our method to be useful in the context of quantum-critical points with inherent dangerously irrelevant operators that cannot 
be tuned away microscopically but whose renormalization flows can be analyzed as we do here for the clock models.
\end{abstract}

\maketitle

The renormalization group (RG) is a powerful framework both for conceptual understanding of phase transitions and for calculations
\cite{Wilson71a,Wilson71b,Fisher72}. A key concept is that a universal critical point can be stable or unstable in the presence of 
perturbations, depending on their scaling dimensions. Similarly, an ordered state can also be stable or unstable under the influence of
perturbations. Under an RG process, a system flows in a space of couplings which change as the length scale is increased under coarse
graining of the microscopic interactions, until finally reaching a fixed point corresponding to a phase or phase transition. At this point,
all the initially present irrelevant couplings have decayed to zero.

RG flows can also be defined of physical observables obtained by Monte Carlo (MC) simulations, allowing controlled finite-size 
scaling analysis---some times referred to as phenomenological renormalization \cite{Fisher72,Binder81,Luck85,Wolff09}. Here we extend the
standard finite-size scaling of a single observable to an entire flow in a space of two observables associated with relevant or irrelevant
couplings. The method is particularly useful for quantifying dangerously irrelevant perturbations (DIPs)---those that are irrelevant at a
critical point but become relevant upon coarse graining inside an adjacent ordered phase \cite{Amit82}. 

{\it Scaling and RG flows.}---Consider a $d$-dimensional lattice model of length $L$ which can be tuned to a critical point by a relevant
field $t$, e.g., the temperature ($t = T_c-T$). With a local operator $m_i$ and its conjugate field $h$, we add $h\sum_i m_i \equiv hM \equiv hL^dm$ 
to the Hamiltonian $H$. In a conventional RG calculation, a flowing field $h'$ is computed under a scale transformation. Here we will instead 
vary the system size, which effectively lowers the energy scale, and calculate the response $\langle m\rangle$ using  MC simulations. Together 
with some quantity $Q$ characterizing the critical point and phases of the system, we can trace out curves (MC RG flows) $(Q,\langle m\rangle)_L$ 
as $L$ increases for fixed values of $h$ and $T$. These flows are very similar to conventional RG flows in the space $(t,h')$.

The singular part of the free-energy density takes the form $f_s(t,h,L)=L^{-d}F_s(tL^{1/\nu},hL^y)$. At $t=0$, the leading $h$ dependent part is 
$f_s \propto hL^{y-d}$, while the statistical mechanics of $H$ gives a contribution $h\langle m\rangle \propto hL^{-\Delta}$ from the internal 
energy. Thus, we obtain the well known relation $y=d-\Delta$. The perturbation is irrelevant at the critical point if $y<0$, but, in the case 
of a DIP, it eventually becomes relevant as $L$ increases in the ordered phase. It has been known for some time that this cross-over is associated 
with a length scale $\xi' \propto t^{-\nu'}$ which may diverge faster than the correlation length $\xi \propto |t|^{-\nu}$ \cite{Oshikawa00}.

To take both divergent length scales properly into account, i.e., to reach the regime where $tL^{1/\nu'}$ is large, we adopt the two-length scaling
hypothesis \cite{Shao16} and write
\begin{equation}
f_s(t,h,L)=L^{-d}F_s(tL^{1/\nu},tL^{1/\nu'},hL^y,\lambda L^{-\omega}),
\label{fscaling}  
\end{equation}  
where we have also included a generic scaling correction with exponent $\omega >0$. The exponents $\nu'$ and $y$ arise from the
same DIP and there is a relationship between them that has been the subject of controversy \cite{Oshikawa00,Lou07,Okubo15,Leonard15}.
Here we will derive the relationship from Eq.~(\ref{fscaling}) and show how the entire RG flow of two observables can be explained.

{\it Models and observables.}---We study three-dimensional (3D) classical clock models on the simple cubic lattice,
\begin{equation}
H=-\sum_{\langle i,j\rangle}\cos(\theta_i-\theta_j)-h\sum_i\cos(q\theta_i),
\label{model}
\end{equation}
with $\theta\in[0,2\pi)$. Based on previous studies \cite{Oshikawa00,Lou07,Okubo15,Leonard15,Hove2003,Hasenbusch11,Pujari15,Ding16},
for $q \ge 4$ the phase transition for fixed $h$ at $T=T_c$ belongs to the 3D U(1) universality class, i.e., the clock field $h$ is irrelevant.
However, for $T<T_c$ it is relevant, reducing the order parameter symmetry from U(1) to $Z_q$ when observed above the DIP length scale $\xi'_q$.

\begin{figure}
\centering
\includegraphics[width=22em]{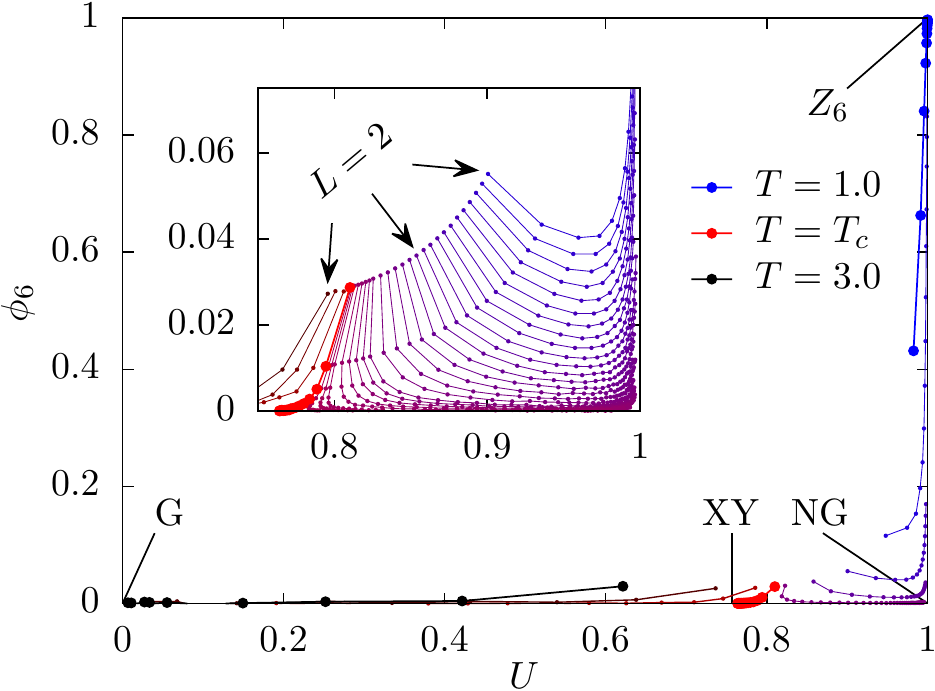}
\caption{MC RG flows for $q=6$. Each set of connected dots represents a fixed $T$ and sizes $L=2,3,4,\ldots$. The sets for the highest and lowest $T$ and $T=T_c$ are shown with bigger dots in black, red
and blue respectively. The inset shows detailed flows in the critical region.}
\label{rg}
\end{figure}

In our MC simulations \cite{Wolff89}, for a given spin configuration we compute $M_x = \sum_{i} \cos(\theta_i)$ and $M_y = \sum_{i} \sin(\theta_i)$.
With  $M = (M_x^2+M_y^2)^{1/2}$ and $\Theta=\arccos(M_x/M)$, an angular order parameter can be defined as
\begin{equation}
\phi_q=\langle \cos(q\Theta) \rangle,
\label{phiqdef}
\end{equation}
which becomes non-zero in response to the $Z_q$ field. This quantity was used to study the length scale $\xi'_q$
\cite{Lou07,Pujari15,Okubo15} (with a slightly different definition in Refs.~\cite{Lou07,Pujari15}), but here we will use it in a different way.
For $T \ge T_c$, $\phi_q \to 0$ when $L \to \infty$, while $\phi_q \to 1$ for $T < T_c$. We will use $\phi_q$ in combination with the
Binder cumulant $U=2-\langle M^4\rangle/\langle M^2\rangle^2$, which takes the limiting forms $U \to 0$ ($T>T_c$), $U\to 1$ ($T<T_c$)
and $U \to U_{\rm XY} \approx 0.757$ (at $T=T_c$ with 3D XY universality \cite{Campostrini06}).

{\it MC RG Flows.}---Fig.~\ref{rg} shows flows of $(U,\phi_q)_L$ for the $q=6$ ''hard'' model, i.e., $h \to \infty$
in Eq.~(\ref{model}). Results for  $q=4,5$ are discussed in Supplemental Material (SM) \cite{sm}, where we also determine $T_c(h)$
for $q=4,5,6$. The RG process is manifested in the flows with increasing $L$ of the two observables at fixed $T$.
The high-$T$ Gaussian fixed point (G) is at $(U,\phi_q)=(0,0)$; the XY critical point at $(U_{\rm XY},0)$, the
U(1) symmetry-breaking Nambu-Goldstone (NG) point at $(1,0)$, and the $Z_q$ symmetry-breaking point at $(1,1)$.
For $T \ge T_c$, we observe simple flows to the fixed points, while for $T<T_c$ there are two stages
in the flow away from the XY point; first toward the NG point and then an NG to $Z_q$ crossover. While this multi-stage
flow is expected based on previous RG results \cite{Oshikawa00,Okubo15,Leonard15}, our description with a phenomenological scaling
function for accessible observables provides a more practical and intuitive framework for numerical simulations.

\begin{figure}
\centering
\includegraphics[width=21em]{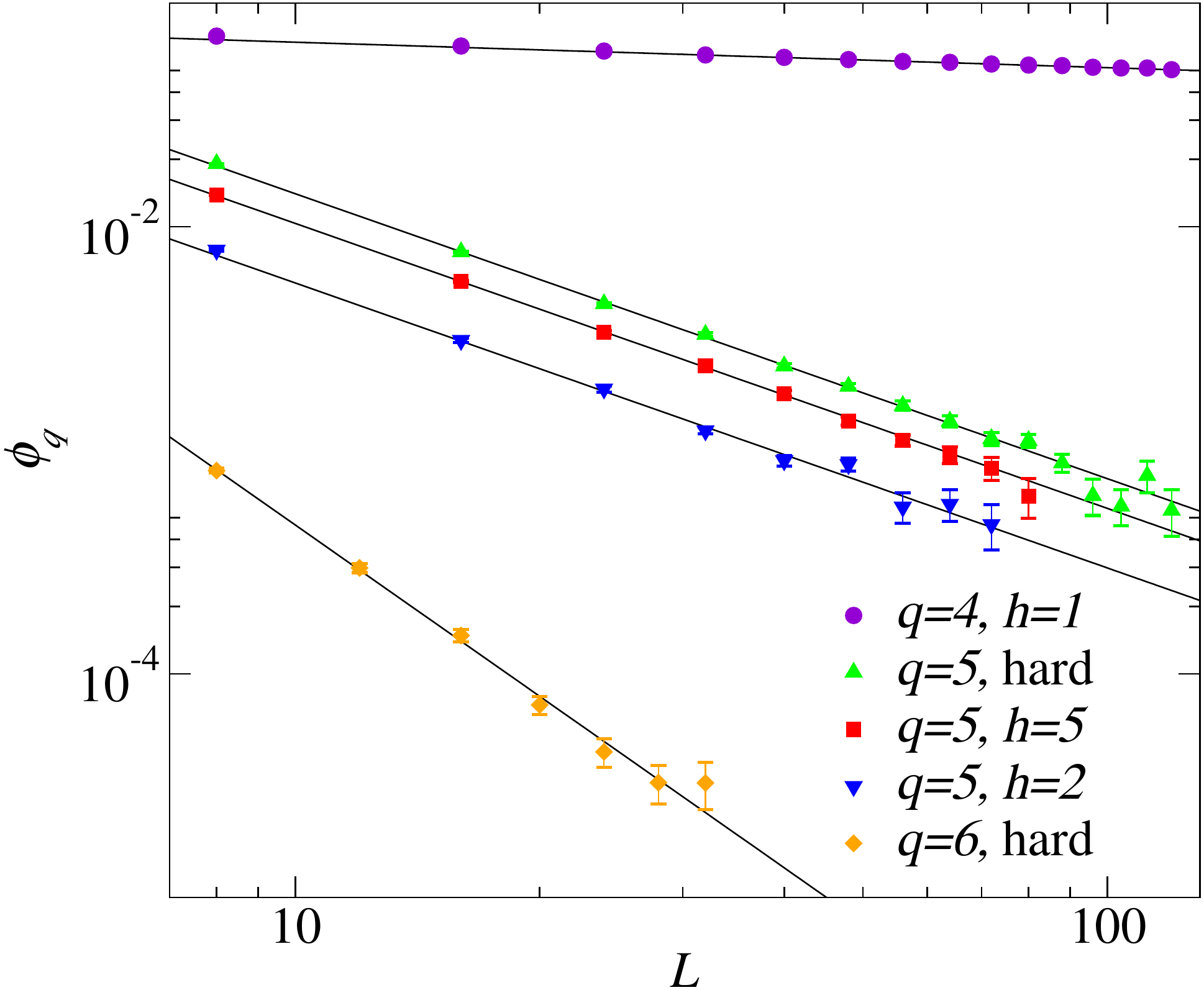}
\caption{Log-log plot of the critical angular order parameter $\phi_q$ vs the linear system size $L$ for several $q$ and $h$ values. 
The fitting lines correspond to the power-law form $\phi_q\propto  L^{-|y_q|}$ and the resulting exponents are summarized in
Table~\ref{yqtable}.}
\label{yqfig}
\end{figure}

{\it Scaling dimensions.}---We first study the scaling dimension $y_q$ of the $Z_q$ field, following the red curve
that tends to the XY fixed point in Fig.~\ref{rg}. Previous MC estimates used $Z_q$ anisotropy correlators in the pure XY model
for $q=4$ \cite{Hasenbusch11}. Since the $Z_q$ field is irrelevant for $q\ge 4$, the decay power $2\Delta_q$ of the correlation
function is larger than $6$, which makes it difficult to determine $\Delta_q$ accurately (see SM \cite{sm} for some results). The decay
of the induced $\phi_q$ is analyzed in Fig.~\ref{yqfig} for $q=4,5,6$ at selected $h$ values. The results listed in Table~\ref{yqtable} 
demonstrate that $\phi_q$ scales as $M=L^dm$ in the general discussion above, i.e., $\phi_q \propto L^{-\Delta_q + d}=L^{-|y_q|}$.

For $q=4$ the $Z_q$ field may only be irrelevant for small $h$; the hard model ($h=\infty$) is equivalent
to two decoupled Ising models, and for $h=2$ the transition already seems to not be in the XY universality class \cite{Pujari15}.
Here we use $h=1$. Our simulations extend up to $L=120$ for $q=4$ but smaller for larger $q$ because of the long runs needed to obtain
sufficiently small error bars on $\phi_q$. To reduce effects of scaling corrections we have excluded small systems until a good fit obtains. Our result
$y_4=-0.114(2)$ agrees well with the best previous numerical result \cite{Hasenbusch11}, but
the error bar is smaller. It also matches a high-order nonperturbative expansion \cite{Leonard15}.
For $q=5$, we have used joint fit to data for several $h$ values, with a common exponent but different prefactors. 
Our result $y_5=-1.27(1)$ is close to an extrapolated value from simulations for smaller $q$ \cite{Okubo15} but differs
significantly from the field-theory expansions \cite{Oshikawa00,Leonard15}. For $q=6$ we obtain $y_6=-2.55(6)$, which again agrees well with the
extrapolated value \cite{Okubo15} but differs from those in Refs.~\cite{Oshikawa00,Leonard15}. For all the $q$ values studied, our results show
that the first-order $\epsilon$-expansion \cite{Oshikawa00} overestimates $y_6$, while the nonperturbative expansion \cite{Leonard15}
underestimates it for $q>4$. All results agree well with a very recent MC calculation of an optimized correlation function \cite{Banerjee18}.

\begin{table}
\centering
\caption{Scaling dimensions $y_q$ of the $Z_q$ field for $q=4,5,6$. The numbers within parenthesis indicate
the statistical errors (one standard deviation) of the preceding digit.}
\begin{tabular}{l|lll}
\hline\hline
\diagbox{$~~~~~~-y_q$}{$q~~~~~~$}     &~~~~~4~~~~~~ &~~~~5~~~~~~ &~~~~6~~~~~~~\\
\hline
~~Ref.~\cite{Oshikawa00}          &~~~ 0.2           &~~~ 1.5           &~~~ 3.0           \\
~~Ref.~\cite{Leonard15}           &~~~ 0.114         &~~~ 1.16          &~~~ 2.29          \\
~~Refs.~\cite{Hasenbusch11,Okubo15}  &~~~ 0.108(6)   &~~~ 1.25          &~~~ 2.5           \\
~~Ref.~\cite{Banerjee18}          &~~~  0.128(6)    &~~~ 1.265(6)       &~~~ 2.509(7)~~~~      \\
~~This work                       &~~~ 0.114(2)      &~~~ 1.27(1)       &~~~ 2.55(6)       \\
\hline\hline
\end{tabular}
\label{yqtable}
\end{table}

Having determined the scaling dimensions, the $Z_q$ order parameter in the ordered phase takes the form
\begin{equation}
\phi_q  = L^{y_q}\Phi(tL^{1/\nu},tL^{1/\nu'_q}),
\label{phiq2}
\end{equation}
where we neglect the irrelevant arguments in Eq.~(\ref{fscaling}) as they merely produce corrections here. We apply this form to 
curves such as those shown in Fig.~\ref{rg}, primarily by defining distances to the various fixed points. We study
$q=6$ specifically but keep the general-$q$ notation.

\begin{figure}
\centering
\includegraphics[width=25em]{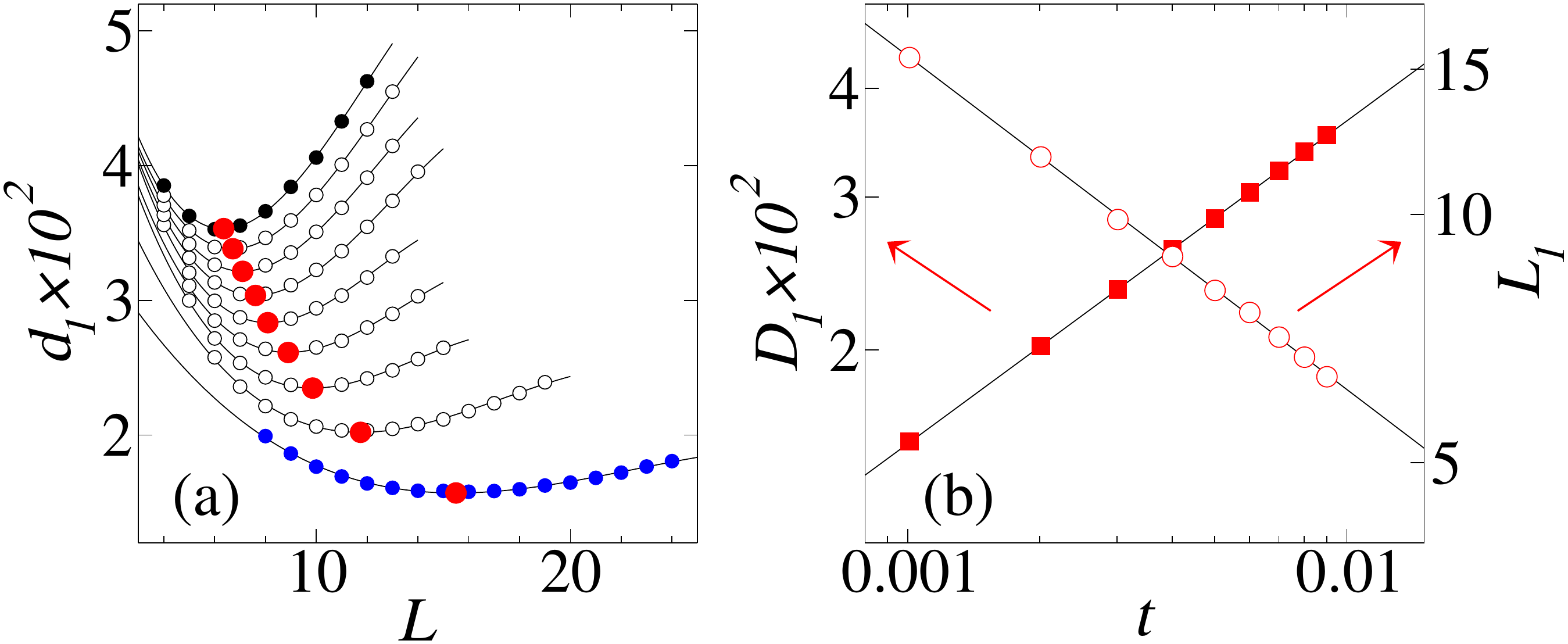}
\caption{(a) Distance $d_1(L)$ to the XY fixed point. Black and blue solid circles correspond to $T=2.193$
and $T=2.201$, respectively, and open circles show temperatures in between. The minimums (red circles) were obtained by polynomial fits. 
(b) Power law behaviors in $t$ of the minimum distance $D_1$ and corresponding size $L_{1}$ [red dots in (a)].}
\label{xy-scaling}
\end{figure}

{\it Scaling near the XY point.}---Though the critical point is well known, it is still useful to study the flows in
the two-dimensional space in Fig.~\ref{rg}. We analyze the minimum distances of the  $T < T_c$ curves to $(U_{\rm XY},0)$. Here 
$tL^{1/\nu'_q} \ll tL^{1/\nu} \ll 1$ in Eq.~(\ref{phiq2}), and to leading order
\begin{equation}
\phi_q \propto L^{y_q}(1+tL^{1/\nu}),
\label{phiq}
\end{equation}
where we do not include unimportant factors for simplicity. The Binder cumulant scales as
\begin{equation}
U=U(tL^{1/\nu}) = U_{XY}+tL^{1/\nu}+L^{-\omega},
\label{u4}
\end{equation}
where $\omega$ is the smallest correction exponent affecting $U$. The scaling form (i.e., without  unimportant factors) of the distance 
$d_1$ to the XY fixed point is
\begin{equation}
d_1 \propto \sqrt{ (tL^{1/\nu}+L^{-\omega})^2 + L^{2y_q}(1+tL^{1/\nu})^2}.
\label{dtl}
\end{equation}
Since $\omega \ll |y_6|$, the first term in the square-root dominates; $d_1 \propto tL^{1/\nu}+L^{-\omega}$, i.e.,
$d_1 \to U-U_{XY}$ here (but not necessarily in general). Minimizing for fixed
$t$ gives the distance $D_1$ and the corresponding system size $L_{1}$
\begin{equation}
D_1 \propto t^{\frac{\omega}{1/\nu+\omega}} = t^{0.345(6)},~~
L_{1}\propto t^{-\frac{1}{1/\nu+\omega}} = t^{-0.440(4)},
\label{dxymin}
\end{equation}
where we have used $\nu=0.6717(1)$ and $\omega =0.785(20)$ \cite{Campostrini06}. Fig.~\ref{xy-scaling}(a) shows $d_1$ versus $L$ and
Fig.~\ref{xy-scaling}(b) shows power-law fits to $D_1(t)$ and $L_{1}(t)$, where the exponents are $0.372(1)$ and $-0.404(4)$, 
respectively. These values are in reasonable agreement with Eq.~(\ref{dxymin}) considering scaling corrections for the rather small sizes
\cite{sm} and the neglected subleading $\phi_6$ contribution in Eq.~(\ref{dtl}). The error bars reflect only statistical fluctuations.

\begin{figure}
\centering
\includegraphics[width=25em]{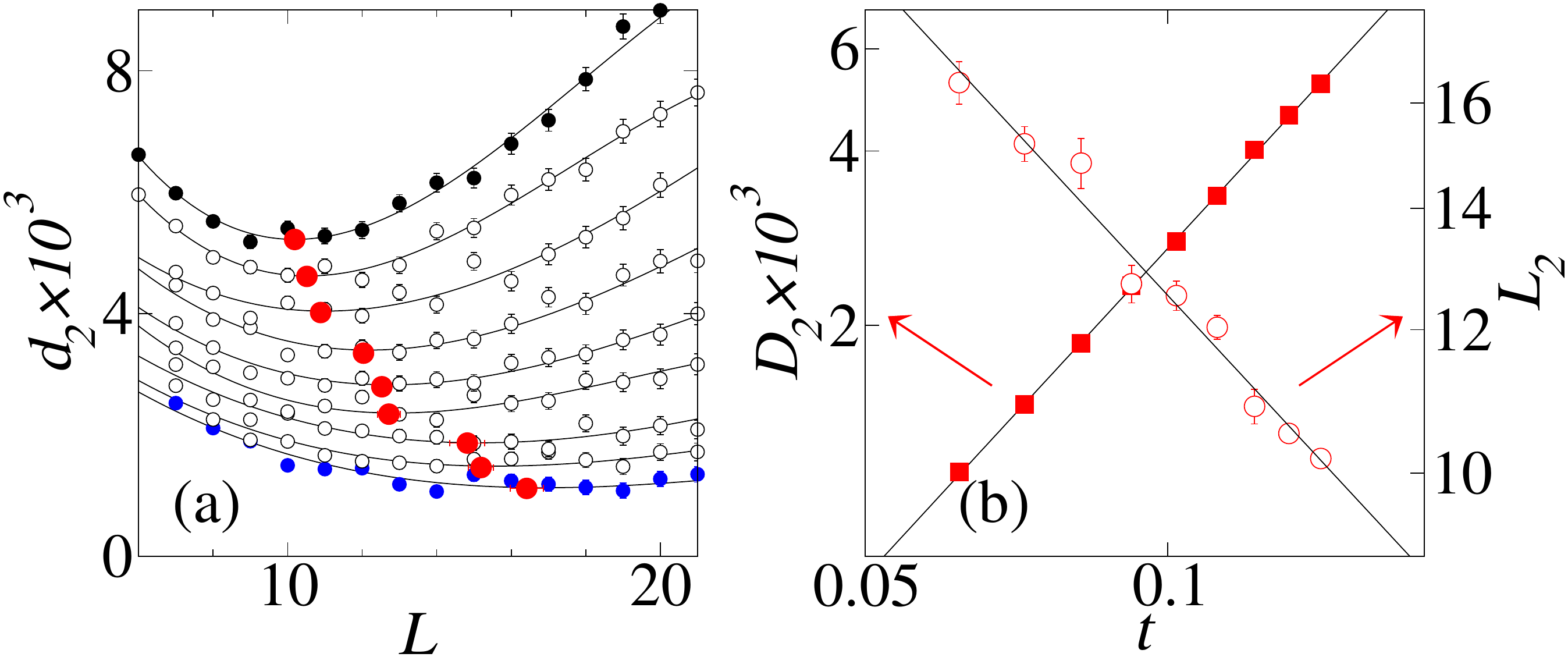}
\caption{(a) Distance of the curves in Fig.~\ref{rg} to the $x$-axis. The black and blue solid circles correspond to $T=2.06$
and $2.14$, respectively, and the open circles are for equally spaced $T$. (b) The minimums (red dots) in (a)
exhibit scaling in $t$ of the minimum distance $D_2$ and the corresponding size $L_{2}$.}
\label{x-scaling}
\end{figure}

Another characteristic of the $T < T_c$ curves in Fig.~\ref{rg} is the minimum distance to the horizontal axis. This RG stage between the XY
and NG fixed points is still governed by the XY criticality because $tL^{1/\nu}$ and $tL^{1/\nu'_q}$ are both small. Since $tL^{1/\nu'_q} \ll tL^{1/\nu}$,
$\phi_q$ is given by Eq.~(\ref{phiq}) and the minimum value $D_2$ and corresponding system size therefore scale with $t$ as (for $q=6$)  
\begin{equation}
D_2 \propto t^{-y_6\nu} = t^{1.71(4)},~~
L_{2}\propto t^{-\nu} = t^{-0.6717(1)}.
\label{dxmin}
\end{equation}
The expected exponents indicated above agree reasonably well with our fits in Fig.~\ref{x-scaling}, where the exponents are $1.88(2)$ and
$-0.60(3)$, respectively. The deviations are again likely due to scaling corrections.

{\it Cross-over exponent $\nu'_q$.---}When $tL^{1/\nu} \gg 1$ but $tL^{1/\nu'_q}$ is arbitrary, Eq.~(\ref{phiq2}) must reduce to
\begin{equation}
\phi_q  = L^{y_q}(tL^{1/\nu})^ag(tL^{1/\nu'_q}),
\label{phiq3}
\end{equation}
where the exponent $a$ follows from the physics of the clock model. Specifically, we can ask how $\phi_q$ depends on $L$ at fixed $t$ when
the $U(1)$ symmetry is barely broken down to $Z_q$, i.e., when $\phi_q \ll 1$. This is a subtle issue at the heart of the long-standing controversy
regarding the symmetry cross-over \cite{Ueno91,Oshikawa00,Lou07,Okubo15,Leonard15}. Instead of invoking physical arguments, we will here simply posit
that $\phi_q \propto L^p$ in the regime where $tL^{1/\nu}$ is large but $tL^{1/\nu'_q}$ remains small [hence $g \approx 1$ in Eq.~(\ref{phiq3})], and later
show how $p$ can be consistently determined from the MC RG flows. Thus, we have $a=\nu(p-y_q)$ in Eq.~(\ref{phiq3});
\begin{equation}
\phi_q  = L^{p}t^{\nu(p-y_q)}g(tL^{1/\nu'_q}).
\label{phiq4}
\end{equation}
This form should apply also when $\phi_q \to 1$, demanding $g  \to (tL^{1/\nu'_q})^b$ with $b=-\nu(p-y_q)$ and $\nu'_q=-b/p$. Then
\begin{equation}
\nu'_q=\nu(1-y_q/p)=\nu(1+|y_q|/p),
\label{nuprime}
\end{equation}
which for $p=3$ agrees with Ref.~\cite{Lou07}, while for $p=2$ it agrees with Refs.~\cite{Okubo15,Leonard15}.
When $\phi_q$ deviates from $1$, $g  \to (tL^{1/\nu'_q})^b[1 - k(tL^{1/\nu'_q})]$, so that for large $tL^{1/\nu'_q}$
\begin{equation}
\phi_q  \to 1-k(tL^{1/\nu'_q}),
\label{phiq5}
\end{equation}
where the function $k$ must be dimensionless.

\begin{figure}
\centering
\includegraphics[width=23em]{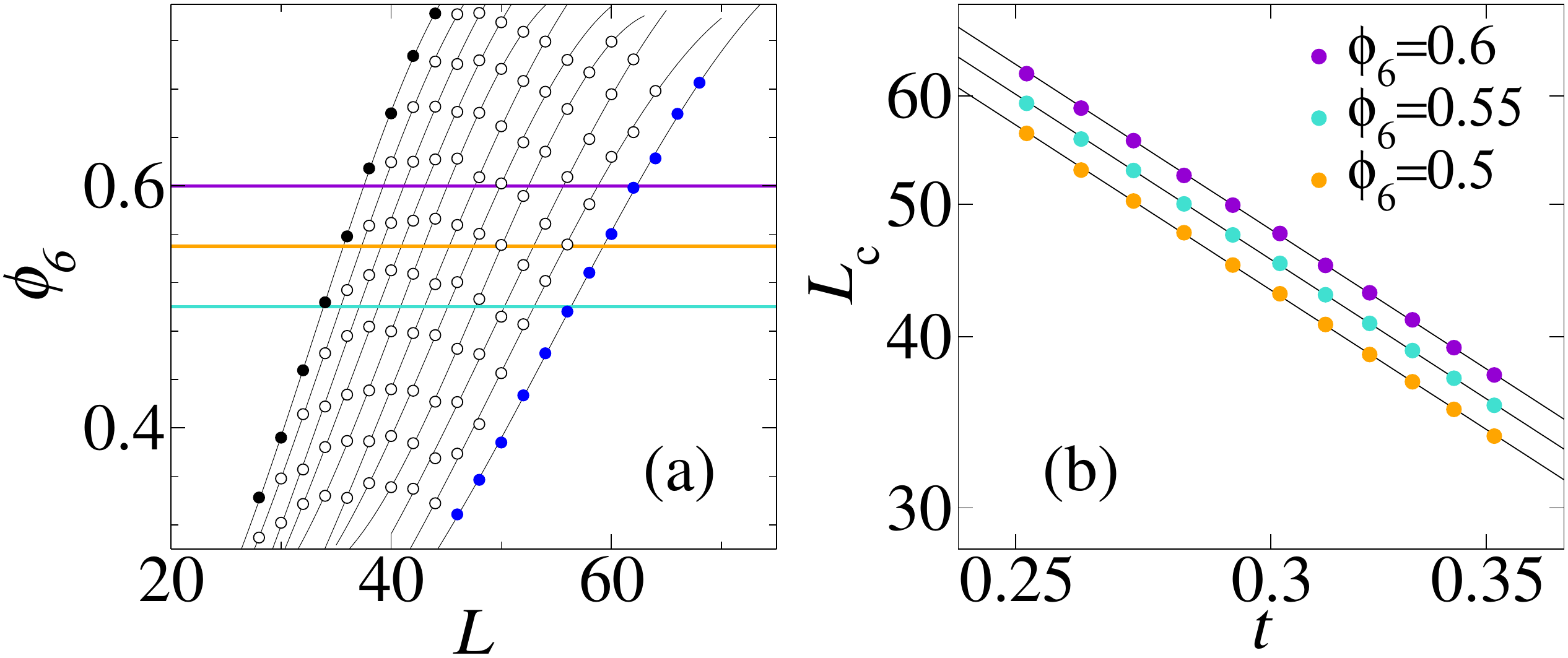}
\caption{(a) $\phi_6$ vs $L$ for temperatures from $T=1.85$ (blue circes) and and $1.95$ (black solid circles).
The crossing points with three horizontal lines at $0.5, 0.55$, and $0.6$ are analyzed in (b) with a joint power-law with a common exponent.}
\label{zq-scaling}
\end{figure}

The exponent $\nu'_q$ in Eq.~(\ref{phiq5}) can be determined by a standard data-collapse procedure \cite{Okubo15,Lou07}. 
Here we proceed in a different way: The function $k(x)$ can be Taylor expanded around some arbitrary point 
$x_0$ where $\phi_q=y_0$; $\phi_q =y_0 + a(x-x_0)$, or $\phi_q=ax + b$ for some $b$. For fixed $t$, we consider
$L=L_c$ for which $\phi_q(L_c)=e$ for some $e$, which gives $L_c \propto t^{-\nu'_q}$. In Fig.~\ref{zq-scaling}(a) we extract $L_c$ for $e=0.5, 0.55$,
and $0.6$. Analyzing the scaling behavior with $t$ in Fig.~\ref{zq-scaling}(b), we find $\nu'_6=1.52(4)$.
Thus, Eq.~(\ref{nuprime}) with $|y_6|=2.55(6)$ is satisfied if $p=2$, in agreement with Refs.~\cite{Okubo15,Leonard15}.
From Eq.~(\ref{phiq4}), the initial growth of $\phi_q$ with $L$ is then $\phi_q \propto L^2$; not $\propto L^3$ \cite{Lou07}.

{\it Near the NG fixed point.}---Finally we consider the distance to the NG fixed point $(1,0)$, where Eq.~(\ref{phiq4}) applies with
$g\approx 1$ ($L \ll \xi'_q$ can be tested self-consistently \cite{sm}). $U$ is close to $1$, but should remain of the form $U(tL^{1/\nu})$
because, as we will see, $L$ and $t$ for a given curve in the region of interest are related such that $t \to 0$ when $L \to \infty$. We need $1-U$,
which has a non-trivial scaling form
\begin{equation}
1- U \propto  (tL^{1/\nu})^{-r},
\label{1mu}
\end{equation}
where it has been argued that, in some cases, $r=d\nu=3\nu$ \cite{Privman84}. However, this result is based on subtle assumptions and
may not be generic \cite{Privman90}. As shown in SM \cite{sm}, $r = 1.52(2) \not=3\nu$ for the XY model.

\begin{figure}[b]
\centering
\includegraphics[width=25em]{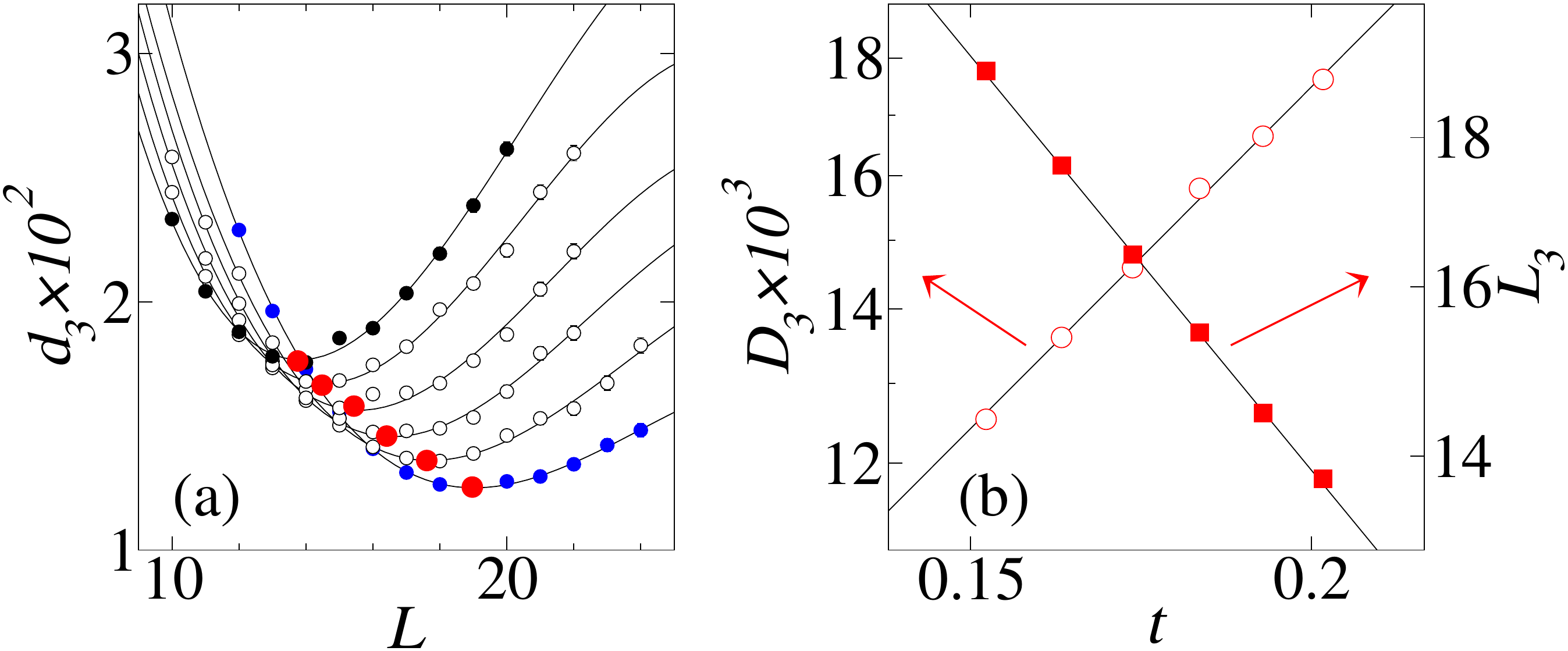}
\caption{(a) The distance $d_3(L)$ to the NG fixed point for temperatures between $T=2.00$ (black solid circles) and $2.05$ (blue circles)
(b) Power-law behaviors in $t$ of the minimum-distance quantities $D_3$ and $L_{3}$.}
\label{ng-scaling}
\end{figure}

The distance to the NG fixed point is, from Eq.~(\ref{1mu}) and Eq.~(\ref{phiq4}) with $\nu(2-y_q)=2\nu'_q$ and $g \approx 1$;
\begin{equation}
d_3 = \sqrt{L^{-2r/\nu}t^{-2r} + L^{4}t^{4\nu_q'}},
\label{d3}
\end{equation} 
and minimizing with respect to $L$ leads to
\begin{equation}
D_{3} \propto \sqrt{t^{2r(R-1)} + t^{4(\nu'_q-R\nu)}},~~
L_{3} \propto t^{-\nu R },~~
\label{ltng}
\end{equation}
where $R= (r+2\nu'_q)/(r+2\nu)$. For the $q=6$ case we then have $D_3  \propto t^{0.9(1)}$ and $L_3  \propto t^{-1.07(3)}$. From the analysis
in Fig.~\ref{ng-scaling} the exponents are $1.19(3)$ and $-1.14(2)$, respectively, in reasonable agreement with the prediction, again considering
that we have not included any scaling corrections. The cross-over behavior around the NG point is also the most intricate of all the regions
in the way the two length scales intermingle.

{\it Discussion.}---The standard finite-size scaling hypothesis in the presence of a DIP (see, e.g., Ref.~\cite{Kenna13}) includes only $tL^{1/\nu}$
and the irrelevant field $hL^y$ in Eq.~(\ref{fscaling}), which is sufficient for extracting the critical exponents close to $T_c$; up 
to $|T-T_c| \propto L^{-1/\nu}$. As we have shown here with the clock model, the other relevant variable $tL^{1/\nu'_q}$ is necessary for describing 
the symmetry cross-over from U(1) to $Z_q$. By considering different necessary (for scaling) limiting forms when the arguments are small or large, we 
have quantitatively explained the entire MC RG flows. 

The controversial relationship between $\nu'_q$ and the scaling dimension $y_q$ \cite{Ueno91,Oshikawa00,Lou07,Okubo15,Leonard15} involves an 
exponent $p$ associated with the initial formation of an effective $Z_q$ symmetric potential for the order parameter. Analytical RG methods for 
related problems, e.g., the Sine-Gordon model with a weak potential are indeed highly non-trivial and sensitive to the type of approximation 
used \cite{Nandori04}. In our approach, $p$ for a given system is obtained from numerical data and can then be used to further understanding 
of the subtle physics of the DIP. We have here confirmed numerically that $p=2$ in the clock model \cite{Okubo15,Leonard15}, but this exponent is 
not necessarily universal---it may depend on a combination of the finite-size properties of the fixed point with the higher symmetry (here 
the well-understood NG point \cite{Hasenfratz90,Dimitrovic91}) and the mechanisms of the DIP causing the lowering of the symmetry.

Our method should be useful in the context of deconfined quantum criticality \cite{Senthil04,Sandvik07,Lou09}, where a scaling ansatz 
with two relevant arguments was introduced to account for anomalous scaling in 2D quantum magnets \cite{Shao16}. There the DIP cannot
be tuned away (unlike some fermionic models \cite{Liu18}), because it is connected to the lattice itself.
Thus, the method introduced here of studying scaling and RG flows in the presence of a  finite DIP is ideal.

\begin{acknowledgments}
 {\it Acknowledgments.}---We would like to thank  Ribhu Kaul, Chengxiang Ding, Jun Takahashi, and Xintian Wu for valuable discussions. 
H.S. was supported by the Fundamental Research Funds for the Central Universities under Grant No.~310421119 and by the NSFC under
Grant No.~11734002. W.G. was supported by NSFC under Grants No.~11734002 and No.~11775021. 
A.W.S was supported by the NSF under Grant No.~DMR-1710170 and by a Simons Investigator Award, and he also gratefully acknowledges
support from Beijing Normal University under YingZhi project No.~C2018046. This research was supported by the Super Computing Center of 
Beijing Normal University and by Boston University's Research Computing Services.
\end{acknowledgments}

\begin{widetext}

\newpage
  
\begin{center}  

\section{Supplementary Information}

{\bf\large Monte Carlo Renormalization Flows in the Space of Relevant and Irrelevant Operators: Application to Three-Dimensional Clock Models}
\vskip5mm

Hui Shao,$^{1,2,*}$ Wenan Guo,$^{3,2,\dagger}$ Anders W. Sandvik,$^{4,5,3,\ddagger}$ 
\vskip3mm

{\it
{$^1$Center for Advanced Quantum Studies, Department of Physics, \\ Beijing Normal University, Beijing 100875, China}\\
{$^2$Beijing Computational Science Research Center, Beijing 100193, China}\\
{$^3$Department of Physics, Beijing Normal University, Beijing 100875, China}\\
{$^4$Department of Physics, Boston University, 590 Commonwealth Avenue, Boston, Massachusetts 02215, USA}}
{$^5$Beijing National Laboratory for Condensed Matter Physics and Institute of Physics,\\ Chinese Academy of Sciences, Beijing 100190, China} \\
\vskip2mm
e-mail: $^*$huishao@bnu.edu.cn, $^\dagger$waguo@bnu.edu.cn, $^\ddagger$sandvik@bu.edu

\end{center}
\vskip5mm

We discuss further results that were used in the main text. In Sec.~1 we determine $T_c$ for
the $q=4,5$ and $6$ clock models. In Sec.~2 we show MC RG flow diagrams for $q=4$ and $5$, complementing the $q=6$ results in Fig.~\ref{rg}
in the main paper. In Sec.~3 we determine the scaling dimensions of the $Z_q$ perturbations using the conventional correlation-function
method for $q=1$-$4$. In Sec.~4 we determine the exponent $r$ governing the asymptotic form of the Binder cumulant $U(x)$ in Eq.~(\ref{1mu}),
by MC calculations for large values of $x=tL^{1/\nu}$ in the ordered phase.
\null\vskip5mm\null

\end{widetext}
\vskip5mm

\setcounter{page}{1}
\setcounter{equation}{0}
\setcounter{figure}{0}
\setcounter{table}{0}
\renewcommand{\theequation}{S\arabic{equation}}
\renewcommand{\thetable}{S\Roman{table}}
\renewcommand{\thefigure}{S\arabic{figure}}

\subsection{1. Determination of critical temperatures}

\begin{figure}[t]
\includegraphics[width=65mm]{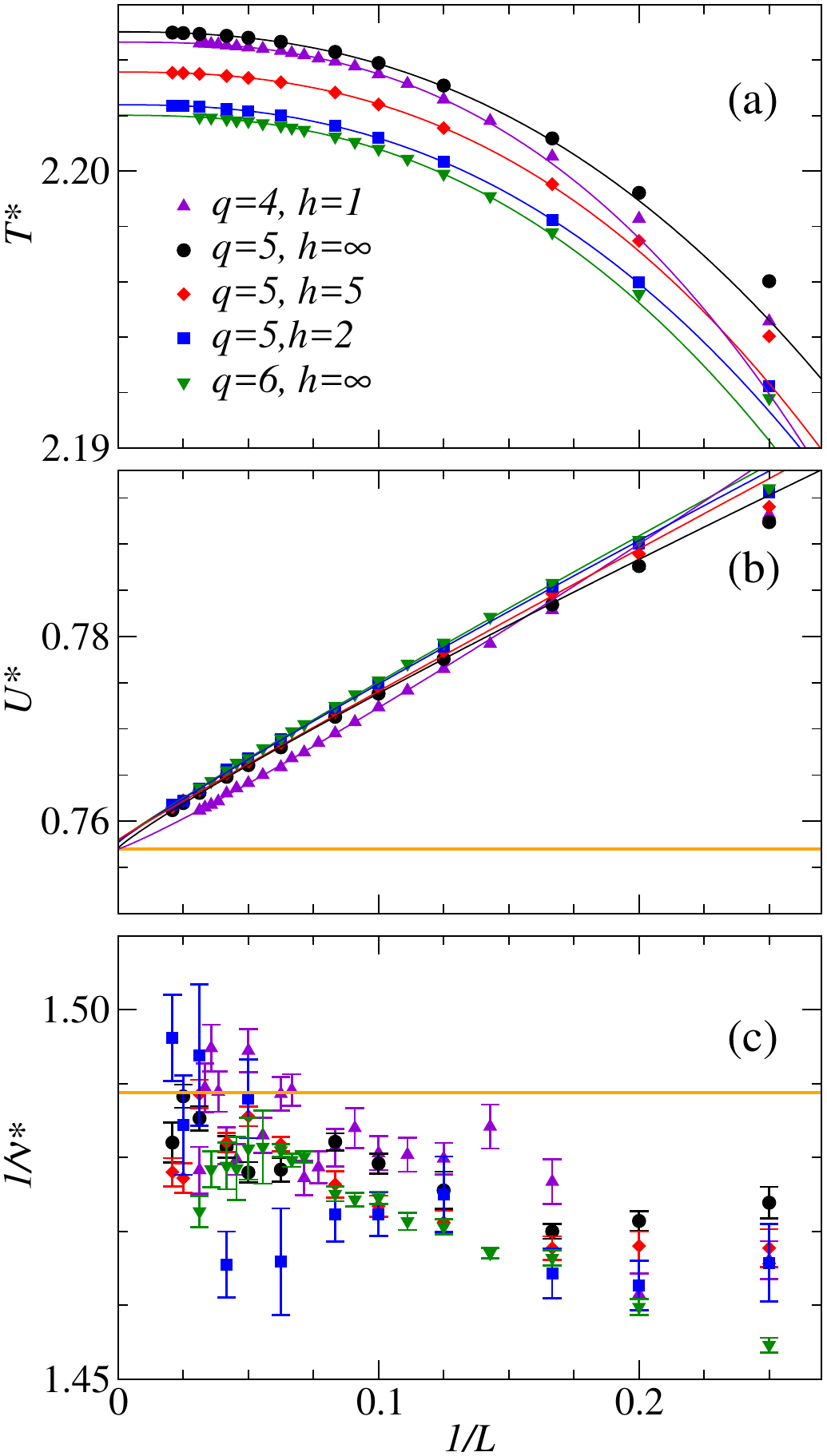}
\caption{Extrapolations of critical-point quantities for different $q$ and $h$ using the $(L,2L)$ cumulant-crossing point analysis 
\cite{Shao16}. The fits to the running critical temperature at which the two cumulants are equal is of the form $T^*(L,2L) = T_c + aL^{-b}$,
where the exponent $b$ should asymptotically (i.e., if sufficiently large system sizes are used) tend to $1/\nu+\omega$. In (b), the cumulant at the
crossing point is scaled as $U^*(L,2L)=U_c+cL^{-e}$, where the exponent $e=\omega$ asymptotically and here we find the effective value $\omega=0.94(3)$.
Small systems were systematically excluded until good fits were obtained. The orange horizontal line shows the expected value of $U_c$ in the 3D O(2)
universality class. In the case of $1/\nu$ in (c), the estimated finite-size values are noisy and we have not carried out fits but merely show consistency 
with the known exponent (horizontal line).}
\label{tcfig}
\end{figure}

To extract the critical temperatures for the clock models with different $q$ and $h$,
we calculate the Binder cumulant of the two-component vector order parameter,
\begin{equation}
U = 2 - \frac{\langle M^4 \rangle}{\langle M^2 \rangle^2},
\end{equation}
with $M=\sqrt{M_x^2+M_y^2}$, where
\begin{equation}
M_x = \sum_{i=1}^{N} \cos(\theta_i), ~~~~~
M_y = \sum_{i=1}^{N} \sin(\theta_i).
\end{equation}
In a standard $(L,2L)$ crossing-point analysis \cite{Luck85} (described in detail and tested, e.g., in the Supplemental Information of
Ref.~\cite{Shao16}), we have computed the cumulant for a series of system sizes around the critical point in each model and used cubic polynomials to
interpolate and extract the crossing points defining the flowing critical temperature $T^*(L,2L)$ and the associated cumulant value $U^*(L,2L)$. In
Fig.~\ref{tcfig}(a,b) we analyze the size dependence of these quantities for all the models studied in the main text.  The infinite-size extrapolated
$T_c$ values are summarized in Table.~\ref{tctable}.
We have also tested the consistency of the critical exponent $\nu$ of the correlation length (obtained from the derivatives $dU/dT$ at the crossing
points) and the universal value of the Binder cumulant $U_c$ with the 3D O(2) universality class \cite{Campostrini06}; the results with increasing $L$
tend to values fully consistent with the known numbers, as shown in Fig.~\ref{tcfig}(b) and (c), though the error bars of the $1/\nu$
estimates are large.

For the $q=6$  case, we also present results for the exponents of the scaling corrections in Fig.~\ref{tcfig}(a) and (b). In (a), the fit
to a power-law correction gives $1/\nu+\omega=2.45(3)$ and in (b) we similarily find $\omega=0.94(3)$ from the correction to $U_c$. These results do
not agree fully with the known values $1/\nu=1.4890(6)$ and $\omega=0.78(2)$ \cite{Campostrini06}, but if we fix $1/\nu$ to its known value
in the estimate for $1/\nu+\omega$, then the value of $\omega$ is statistically consistent with the value from the $U^*$ fit. Since we have
only included one correction here, and influence from the higher-order corrections may be significant still at these system sizes, the exponent
$\omega$ should be considered as an ``effective exponent'', whose value should approach the true value for system sizes larger than those used here.

\begin{table}
\centering
\caption{Critical temperatures for various $q$ and $h$ values. The underlying analysis is presented in Fig.~\ref{tcfig}.}
\begin{tabular}{l|ccc}
\hline\hline
 \diagbox{$~~~~~h$}{$q~~~~$}&               4&               5&               6\\
 \hline
 ~~~~1.0~~~~                &~~~2.20465(1)~~~&                &                \\
 ~~~~2.0~~~~                &                &~~~2.20239(1)~~~&                \\
 ~~~~5.0~~~~                &~~~          ~~~&~~~2.20357(1)~~~&~~~          ~~~\\
 ~~~~$\infty$~~~~           &                &~~~2.20502(1)~~~&~~~2.20201(1)~~~\\
\hline\hline
\end{tabular}
\label{tctable}
\end{table}

\subsection{2. MC RG Flows for the $q=4,5$ clock models.}

\begin{figure}
\includegraphics[width=70mm]{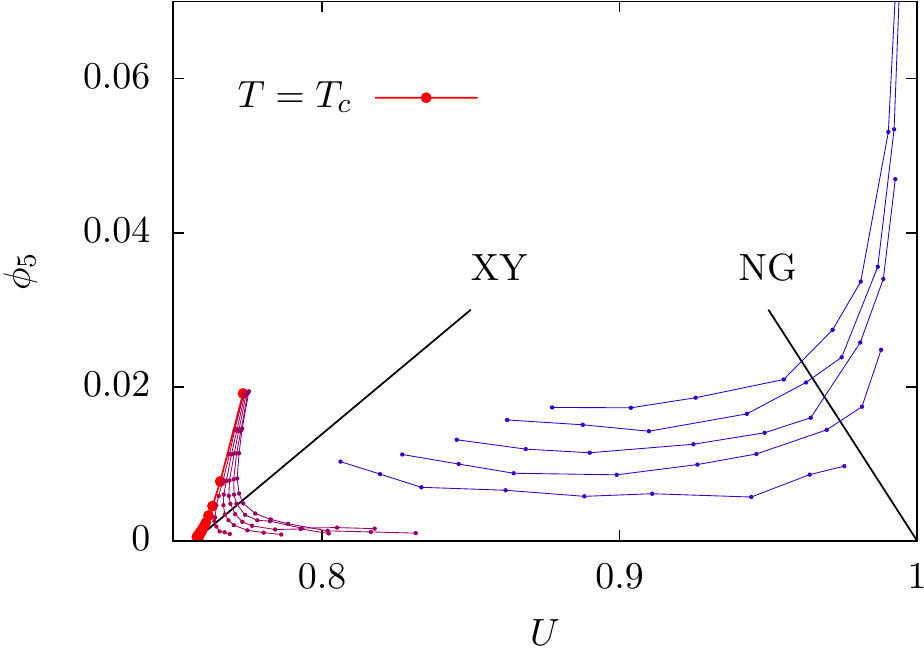}
\caption{MC based RG flows for the $q=5$ hard clock model. Each set of connected dots represents a fixed $T$ and different system sizes.
The flows for increasing $L$ are directed toward the edges of the graph. We show one set of curves close to the critical flow (shown in red,
with the larger points corresponding to $T=T_c$), and another set (shown in blue) at lower temperatures where the cross-over to the $Z_5$
point at $(U,\phi_6)=(1,1)$ can be observed clearly. The system sizes start at $L=8$ for the group of curves at and close to $T_c$ and at
$L=16$ for the other group. The maximum sizes vary from $L=48$ to $120$. Statistical errors are reflected in the degree of unsmoothness
in the curves.}
\label{rg-5}
\end{figure}

In addition to the $q=6$ MC RG flows discussed in the main paper, we have performed more limited simulations of the cases $q=4$ and $5$.
Results for the $q=5$ hard-constrained model is shown in Fig.~\ref{rg-5}, with data distributed mostly near the XY and NG fixed points.
For the system sizes available, there is no $T$ for which we can observe both the flow toward the NG fixed point and the cross-over away
from this point toward the $Z_5$ fixed point. However, we can see these parts of the flows separately for suitably chosen temperatures
(the two groups of curves in Fig.~\ref{rg-5}). On a qualitative level the flows are very similar to the $q=6$ case.

\begin{figure}
\centering{}
\includegraphics[width=70mm]{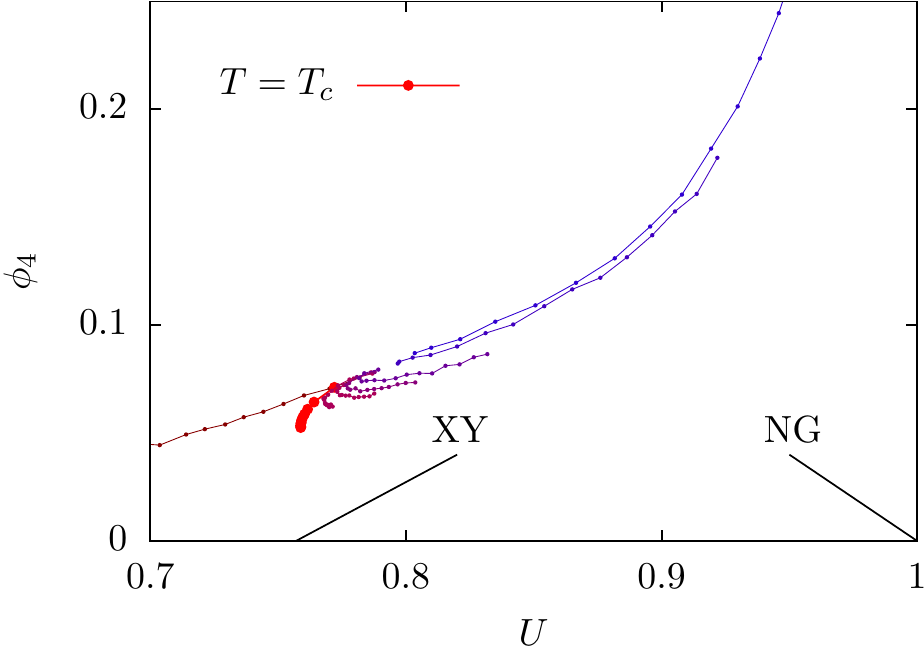}
\caption{MC based RG flows for the $q=4$ clock model with $h=1$. Each set of connected dots represents a fixed $T$, with the dots
corresponding to system sizes $L = 4,6,8,\ldots,32$. The $T = T_c$ points are shown with bigger red dots corresponding to system sizes
$L=8,16,24,\ldots,120$.}
\label{rg-4}
\end{figure}

Figure \ref{rg-4} shows results for  the case $q=4$, $h=1$. At first sight, the flows here appear to be very different from the $q=5,6$ cases.
However, this should just be due to the small scaling dimension of the $Z_4$ field, $|y_4| \approx 0.11$ (Table I in the main paper). This means
that $\phi_4$ decays very slowly with increasing system size, as is clear both from Fig.~2 in the main paper and the red set of $T_c$ data in
Fig.~\ref{rg-4}. For the system sizes available, there are not yet any sign of flows toward the NG point before the ultimate flow toward the $Z_4$ fixed point.
For very large system sizes we expect that such cross-over behavior should be manifested also in this case, but to
observe it requires a clear separation of the length scales $\xi \propto t^{-\nu}$ and $\xi'_q \propto t^{-\nu'_q}$. Since in this case the
difference between the exponents is very small, $\nu'_4-\nu = \nu |y_4|/2 \approx 0.04$, if we would like to have, say, $\xi'/\xi = 10$, we need 
$t \approx 10^{-25}$ (assuming all proportionality factors are of order one). From our analysis of the flow away from the NG fixed point, summarized
as Eq.~(\ref{ltng}), we then have roughly $L_c \propto t^{-\nu R} \approx 10^{18}$ for the system size where the cross-over will occur. This length scale
is clearly beyond any current or future MC calculations.

It is also important to check whether it is always true that $tL^{1/\nu'}$ asymptotically vanishes when the cross-over in the neighborhood of
the NG point takes place. This was the assumption under which we derived the cross-over point with minimum distance $D_3$ to the NG point and
the associated length $L_3$, because we set $g=1$ in the scaling form Eq.~(\ref{phiq3}) of $\phi_q$. Rewriting the scaling form of $L_3$ in
Eq.~(\ref{ltng}) as
\begin{equation}
L_3 \propto t^{-\nu R}=t^{-\nu(r+2\nu'_q)/(r+2\nu)},
\end{equation}
we have that $t$ scales with $L_3$ as
\begin{equation}
t \propto L_3^{-(r/\nu+2)/(r+2\nu'_q)}.
\end{equation}
Thus, the relevant scaling argument corresponding to the second length scale depends on $L_3$ as
\begin{eqnarray}
tL_3^{1/\nu'_q}&\propto& L_3^{1/\nu'_q-(r/\nu+2)/(r+2\nu'_q)}\\
              &=&L_3^{\frac{r(1-\nu'_q/\nu)}{\nu'_q(r+2\nu'_q)}},
\end{eqnarray}
where that the exponent on $L_3$ is always negative because $\nu'_q>\nu$ for a DIP. Therefore,
\begin{equation}
g(tL_3^{1/\nu'_q})  \to g(0) = 1,
\end{equation}
and the self-consistency of the assumption is confirmed for any $q\ge 4$ in the neighborhood of the NG cross-over.

\subsection{3. Scaling dimensions $y_q$ from correlation functions in the XY model}

The standard way to obtain the scaling dimension of an irrelevant or relevant operator is to compute the related correlation
function at the critical point in the model without the perturbation. In the case of the 3D XY model, the best MC calculation
of the scaling dimension of the $Z_4$ clock perturbation is in Ref.~\cite{Hasenbusch11}. Because of the rapid decay of the
correlation functions for larger $q$, no MC results based on the conventional method are available for $q>4$, as far as we
are aware. Our method presented in the main paper can reach larger $q$ because of the slower decay of the induced operator
expectation value in the presence of the perturbation.

Here we contrast the conventional and new method by considering the $q=4$ case, computing the $Z_q$ correlator with MC
simulations at the 3D XY  critical point,  using $T_c=2.20184$ \cite{Campostrini06}. The local operator corresponding to the
$Z_q$ field can be taken as:
\begin{equation}
m(q,r_i)=\cos(q\theta_i),
\end{equation}
and we study the corresponding correlation function
\begin{equation}
C(q,r)=\langle m(q,r_i)m(q,r_j) \rangle=\langle \cos(q\theta_i-q\theta_j)\rangle,
\label{cqr1}
\end{equation}
where $r=r_i-r_j$ and the global rotational symmetry has been taken into consideration. 

\begin{figure}
\includegraphics[width=70mm]{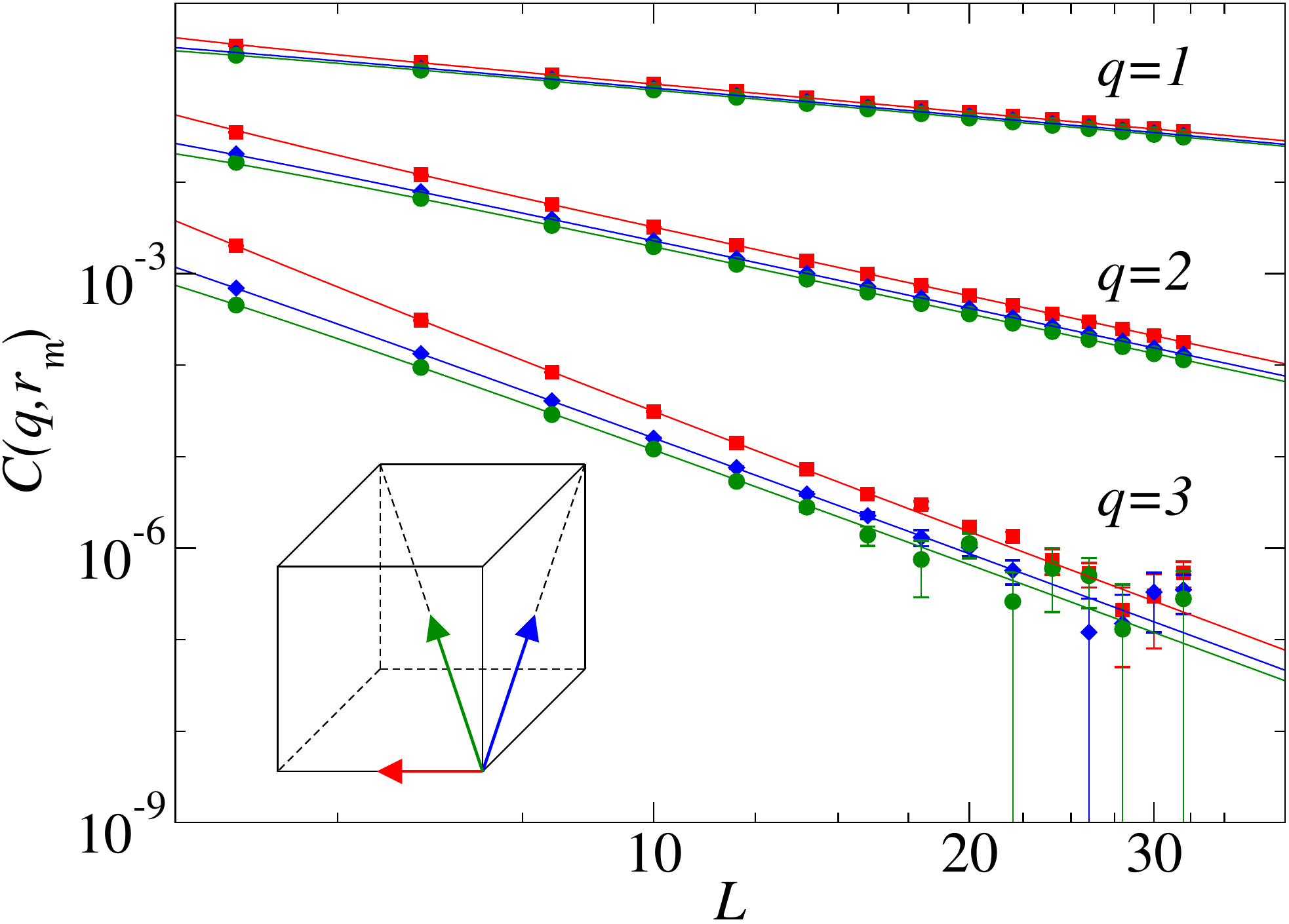}
\caption{The correlation function $C(q,r_{\text m})$ defined in Eq.~(\ref{cqr1}) vs $L$ for $q=1,2,3$ (relevant perturbations). 
Joint fits along all three directions were performed according to Eq.~(\ref{cqr2}). For given $q$, we impose the same decay exponent for all three
directions as well as a common exponent of a scaling correction. The resulting scaling dimensions are listed in Table \ref{xy-exponent}.}
\label{xy}
\end{figure}

In Fig.~\ref{xy} we analyze the long-distance correlation function $C(q,r_{\text m})$ in the three different lattice directions
(i.e., $r_{\text m}$ is half the system length in the respective directions), as indicated in the inset of the figure. The asymptotic
form should be
\begin{equation}
C(q,r_{\text m})\sim aL^{-2\Delta_q}(1 + bL^{-\omega}),
\label{cqr2}
\end{equation}
where $\Delta_q=3-y_q$, with $y_q$ being the scaling dimension of the $Z_q$ field, and we have also included
a scaling correction with exponent $\omega$. We perform joint fit to Eq.~(\ref{cqr2}) with the MC data along all three directions,
where same exponents but different prefactors $a$ are used.

In Fig.~\ref{xy} we present results for $q=1,2,3$, i.e., the cases in which the $Z_q$ fields are relevant. The results for the scaling
dimensions are summarized in Table \ref{xy-exponent} and compared with previous MC studies \cite{Campostrini06,Hasenbusch11}. The agreement
is good, and in the case of $q=2$ we improve on the statistical error. We should note here that the previous study used a system-volume
integrated correlator, for which the statistical errors of the correlations are smaller but the corrections may be larger.

\begin{table}
\centering
\caption{Scaling dimension of the $Z_q$ field based on the fits in Fig.~\ref{xy} and compared with previous numerical results.}
\begin{tabular}{c|ccc}
\hline\hline
~~~~$q$~~~~&~~~~~~~~~~1~~~~~~~~~~&~~~~~~~~~~2~~~~~~~~~~&~~~~~~~~~~3~~~~~~~~~~\\
\hline
~~~$y_q$  &2.481(1) &1.7677(4) &0.876(13)\\
          &2.4810(3) \cite{Campostrini06} &1.7639(11) \cite{Hasenbusch11} &0.8915(20) \cite{Hasenbusch11}\\
\hline\hline
\end{tabular}
\label{xy-exponent}
\end{table}

\begin{figure}[t]
\includegraphics[width=70mm, clip]{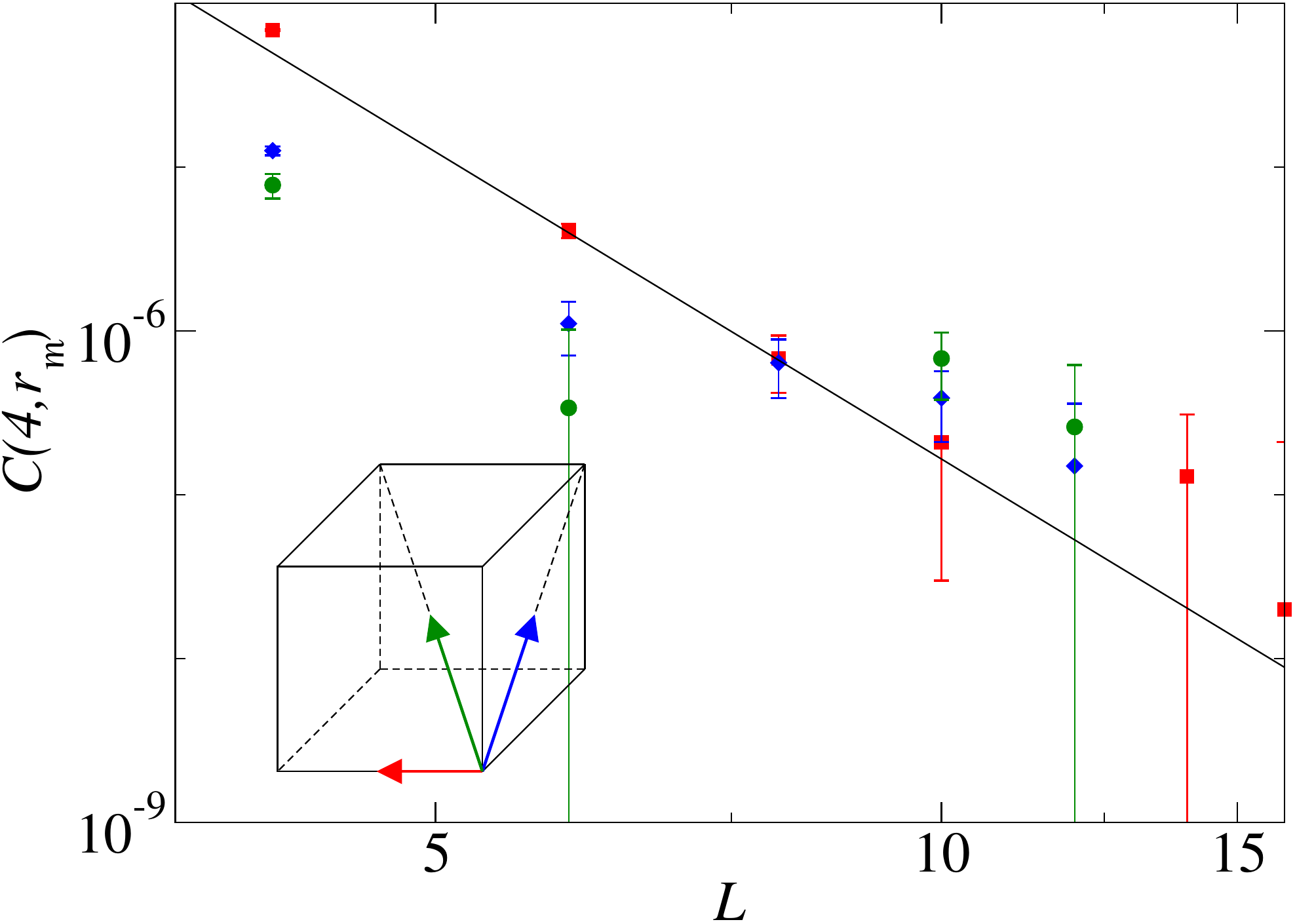}
\caption{The $Z_q$ anisotropy correlation function for $q=4$. The line has slope given by the scaling dimension
$-2\Delta_4=-6+2y_4 = -6.23$ from Table \ref{yqtable} in the main paper.}
\label{xy-2}
\end{figure}

When the $Z_q$ field becomes irrelevant, the decay exponent of the correlation function grows larger than $6$, 
and it becomes extremely hard to extract the scaling dimension in this way. We show our $q=4$ data in Fig.~\ref{xy-2}.
Here we do not report any results of fitting, but only indicate the expected decay power $2\Delta=(3+|y_4|) \approx 6.23$
based on the scaling dimension $y_q \approx -0.114$ extracted with the alternative method in the main paper.

Here we should again note that the previous MC study \cite{Hasenbusch11} used a system-integrated correlator, for which the decay exponent is $2(3-\Delta_q)=2y_q$. 
With the larger exponent due to summation over the system volume, the error bars are significantly reduced and the results were therefore
considerably less noisy than in the data presented here. The long-distance correlator is possibly less affected by scaling corrections,
though we have not tested this. Our approach of explicitly including the field still appears to work better, having a decay exponent of just
$y_q$. Our main purpose of studying the $Z_q$ correlation functions here was mainly to  establish the consistency between the two approaches.

\subsection{4. Asymptotic form of the Binder cumulant}

Recall that, in the critical finite-size scaling form of some singular quantity $A$,
\begin{equation}
A(t,L)=L^{\sigma}g(tL^{1/\nu}),
\label{atl}
\end{equation}  
the exponent $\sigma$ must be compatible with the asymptotic form of the scaling function $g(x)$, $x=tL^{1/\nu}$. This behavior is
connected to the size-independent scaling form in the thermodynamic limit, $A \propto t^\kappa$ (where $\kappa$ is a generic notation for
the critical exponent for the quantity in question), which is obtained if $g \to x^\kappa$ when $x\to \infty$ (i.e., $L \to \infty$ for
fixed small $t$). Then, to eliminate the $L$ dependence we must have $\sigma=-\kappa/\nu$.

In the case of the dimensionless Binder cumulant $U$, $\sigma=0$ and, accordingly, the corresponding scaling function $g(x)$ in Eq.~(\ref{atl})
must take the form $g \to c$, where $c$ is a constant which we know takes the value $c=1$ in the ordered phase (while $c=0$ in the
disordered phase). The scaling form does not immediately tell us how $g$ approaches $1$, however, which is what we need in the analysis
of the flow close to the NG fixed point in the main paper. It should be noted that the scaling regime of interest here does not yet correspond to
Gaussian fluctuations in the ordered phase, because $t$ approaches zero with increasing length-scale, as shown in the main paper.
A natural assumption is that $1-g(x)$ takes a power-law form, $1-g(x) \propto x^{-r}$, corresponding to the form of $1-U$ in Eq.~(\ref{1mu}).
The exponent $r$ should presumably also be related to the critical exponents of the universality class in question.

Surprisingly, while the Binder cumulant is one of the most important quantities used to characterize critical points in numerical
studies \cite{Binder81,Luck85}, the asymptotic form of $1-U$ has not been extensively studied---the focus has naturally been on the behavior for small
arguments; $x=0$ and $x \approx 0$. We are only aware of Privman's work on the asymptotic $x \to \infty$ behavior \cite{Privman84,Privman90}.
He argued that $r=d\nu$ but also pointed out that the assumptions underlying this conclusion are somewhat speculative and untested.

\begin{figure}
\includegraphics[width=70mm, clip]{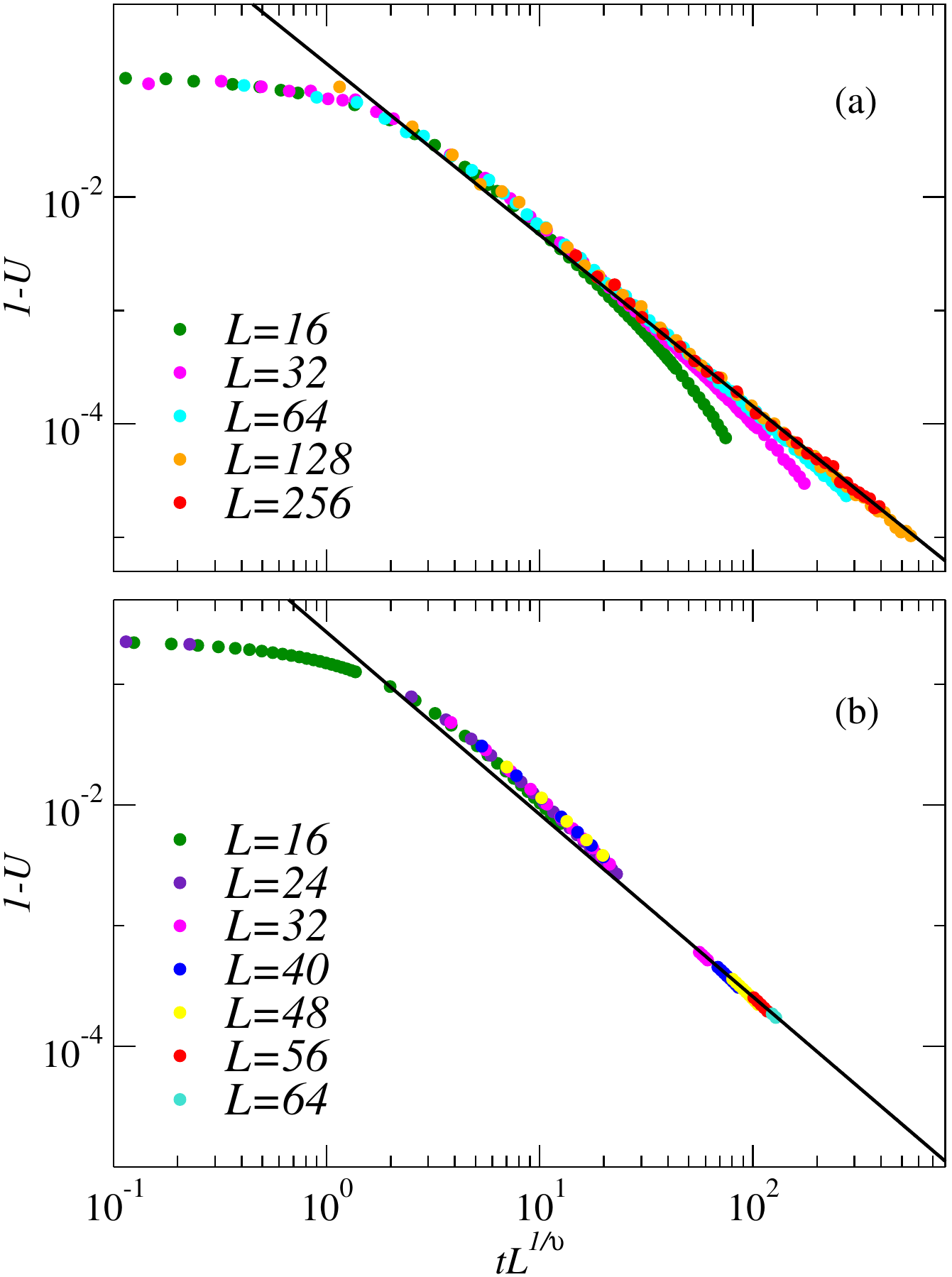}
\caption{MC results for the $1-U$ of the 3D XY model (a) and the $q=6$ hard clock model (b) for different system lengths $L$.
The temperatures are below $T_c$ for each model and the data are shown versus $tL^{1/\nu}$, with $t=T_c-T$. In (a), the line is a fit
to the $L=256$ data, giving the exponent $r=1.52(2)$. In (b), the line has the same slope and is drawn through the data
sets for $tL^{1/\nu}$ in the range $50 \sim 100$.}
\label{u1fig}
\end{figure}

To investigate the scaling behavior, we have carried out systematic MC calculations of the 3D XY model and the $q=6$ clock model inside their ordered
phase in order to extract the exponent $r$ independently.
Our results for the XY model are shown in Fig.~\ref{u1fig}(a). We performed dedicated simulations targeting $1-U$
for system sizes up to $L=256$ for a wide range of the scaling variable $tL^{1/\nu}$, sufficient to reliably observe data collapse and
an asymptotic power-law form. A fit to the $L=256$ data gives the exponent $r=1.52(2)$, which is clearly different from Privman's prediction
$r=3\nu \approx 2.02$ \cite{Privman84,Privman90}. As Privman pointed out, there are subtle assumptions made in the derivation of his
result, and the behavior may not be generic. In the case here, the exponent is consistent within statistical errors with the exponent
$1/\nu$, but we see no obvious reason for this value.

In the case of the clock model, results for which are shown in Fig.~\ref{u1fig}(b),
we have just plotted the same data that we used in the main paper, going up only to $L=64$. The data forming a group
in the range $tL^{1/\nu} \approx 50 \sim 100$ are fully consistent with the same exponent as in the XY model, and for lower values of the
scaling variable the behaviors are also very similar. For small and moderate values of $tL^{1/\nu}$ it is clear that the clock and XY models
should behave very similarly in this regard, since the clock field close to $T_c$ is irrelevant. However, when $tL^{1/\nu}$ is larger, e.g.,
when $tL^{1/\nu} \approx 100$ in in Fig.~\ref{u1fig}(b), there could in principle be a cross-over behavior also in $U$, where $tL^{1/\nu'_q}$
may impact the scaling behavior (perhaps as a correction) when it also reaches large values. We do not see any evidence of a break-down of
the $tL^{1/\nu}$ scaling, however.

It would be interesting to study $1-U$ also for other models, to test the generality of the results found here.

\end{document}